\documentclass[pdflatex,sn-mathphys, iicol]{sn-jnl}

\jyear{2022}%

\theoremstyle{thmstyleone}%
\theoremstyle{thmstyletwo}%

\theoremstyle{thmstylethree}%

\usepackage{soul}
\usepackage{natbib}

\begin{document}

\title[Article Title]{Quantitative analysis of dynamic CT imaging of methane-hydrate formation with a hybrid machine learning approach}


\author*[1,3]{\fnm{Mikhail I.} \sur{Fokin}}\email{fokinmi@ipgg.sbras.ru}

\author[2]{\fnm{Viktor V.} \sur{Nikitin}}\email{vnikitin@anl.gov}
\equalcont{These authors contributed equally to this work.}

\author[1,3]{\fnm{Anton A.} \sur{Duchkov}}\email{duchkovaa@ipgg.sbras.ru}
\equalcont{These authors contributed equally to this work.}

\affil*[1]{\orgname{Institute of Petroleum Geology and Geophysics SB RAS}, \orgaddress{\street{3 Ac. Koptyuga Avenue}, \city{Novosibirsk}, \postcode{630090}, \country{Russia}}}

\affil[2]{\orgdiv{Advanced Photon Source}, \orgname{Argonne National Laboratory}, \orgaddress{\street{9700 S. Cass Avenue}, \city{Lemont}, \postcode{60439}, \state{IL}, \country{USA}}}

\affil[3]{\orgname{Novosibirsk State University}, \orgaddress{\street{1 St. Pirogova}, \city{Novosibirsk}, \postcode{630090}, \country{Russia}}}


\abstract{Fast multi-phase processes in methane hydrate-bearing samples are challenging for micro-CT quantitative study because of complex tomographic data analysis involving time-consuming segmentation procedures. This is due to the sample multi-scale structure changing in time, low X-ray attenuation and phase contrast between solid and fluid materials, as well as large amount of data acquired during dynamic processes. We propose a hybrid approach for automatic segmentation of tomographic data from time-resolved imaging of methane gas-hydrate formation in sandy granular media. First, we use an optimized 3D U-net neural network to perform segmentation of mineral grains that are characterized by low contrast to the surrounding pore brine-saturated phases. Then, we perform statistical clustering based on the Gaussian mixture model for separating the pore-space phases that are characterized by gray-level instabilities caused by dynamic processes during hydrate formation. The proposed approach was used for segmenting several hundred gigabytes of data acquired during an in-situ tomographic experiment at a synchrotron. Automatic segmentation allowed for studying properties of the hydrate growth in pores, as well as dynamic processes such as the incremental pore-brine flow and redistribution.}




\maketitle

\section{Introduction}\label{intro}

Natural gas hydrates are icy-looking crystals formed from water and methane. They accumulate worldwide in favorable thermobaric conditions (in permafrost regions on land and in bottom sediments offshore) forming an important alternative source of methane \cite{birchwood2010developments}.  Dissociation of natural gas hydrates may affect global climate change due to associated greenhouse gas emissions \cite{ruppel2017interaction}, cause a number of geo-hazards including instability of wells and infrastructure, landslides and sinkholes in the permafrost regions \cite{koh2007natural,yakushev2000natural}. Another use of gas hydrates is related to the problem of geological carbon sequestration, which is important for reducing anthropogenic gases safely and efficiently \cite{lackner2003guide}. In particular, depleted reservoirs can be used for injecting the CO$_2$ and storing it in the form of hydrates \cite{park2006sequestering,koide1997deep}. 
Thus there is an ongoing activity of exploration and characterization of natural gas-hydrate accumulations by geophysical methods  \cite{spence2010geophysical}.
Laboratory experiments with environmental cells are widely used for forming gas hydrates in porous rock samples and studying related changes in their physical properties \cite{waite2009physical,wang2021wave,dugarov2021acoustic}. Dynamic imaging at micro scale may further improve our understanding of how gas-hydrate formation/dissociation in geomaterials depends on the rock type, pore structure, and other factors. 

X-ray computed tomography (CT) is widely used for high-resolution non-destructive imaging across science and engineering applications \cite{withers2021x}, including Geosciences \cite{mees2003applications}. In-situ CT imaging is of special importance for studying geomaterials as it helps to mimic thermo-baric conditions typical for the Earth interior \cite{fusseis2014low}. 
The use of synchrotron radiation sources allows for time-resolved CT imaging of fast processes in geomaterials, including geomechanical deformation and failure of rocks \cite{zhang2022dynamic}, gas-hydrate formation and dissociation in porous samples \cite{nikitin2020dynamic,nikitin2021dynamic}, dynamic fluid flow through pore pathways \cite{dobson20164}, and other pore-scale processes.  

Fast data acquisition rates in synchrotron CT experiments yield a huge amount of stored data that need to be analyzed. In some cases, experimental results are virtually impossible to store on hard drives and different data reduction algorithms have to be applied already during tomographic experiments. In particular, one may need fast procedures for feature extraction and tracking changes \cite{tekawade20213d}. Accurate 3D volume segmentation is the most important and time-consuming step in synchrotron data analysis.
In many situations, low data contrast and artefacts in reconstructions complicate automatic segmentation procedures, causing users to switch to ultra-slow  manual segmentation instead of testing more optimal methods and improving reconstruction quality.        

Fast segmentation procedures also help in localizing regions of interest inside the sample and selecting time intervals for studying dynamic phenomena in detail.
Finally, segmentation allows for quantitative estimation of components in a sample and forms the main input for Digital Rock Physics where numerical simulations are used to predict macro-properties of natural rock samples \cite{wang2015permeability,  sell2016path, zhang2020pore}.
Image-computed rock properties may vary considerably depending on segmentation results \cite{saxena2017effect,andra2013digital1,andra2013digital2,rezaei2019effectiveness}, affecting the accuracy of numerical simulations.

Micro-CT images can be segmented by using various methods including global thresholding \cite{iassonov2009segmentation}, marker-based watershed algorithm \cite{zhang2012improvement}, and clustering algorithms \cite{kang2009comparative}. However, one of the main difficulties in studying methane hydrate-bearing media by X-ray CT is a weak contrast between pore water and methane hydrate crystals (due to similar densities). Although some researchers use heavier gas (Xenon) for the gas-hydrate formation to enhance  contrast levels between different materials \cite{chaouachi2015microstructural, yang2016synchrotron}, physical properties become different from the desired methane gas hydrate \cite{fu2019xenon}. Alternatively, pore brine can be formed with heavier salts  (NaBr, KI) instead of natural NaCl \cite{lei2018laboratory}. 
This method is more suitable for modeling gas hydrates in porous media, and it provides good contrast between the brine and hydrate. However, it also results in another low-contrast issue -- now between the pore brine and grains of the mineral matrix.

Thus, there is a need for better segmentation techniques for the automatic analysis of images acquired in gas-hydrate studies. 
The global thresholding method in the case of images with close gray-level averages can be improved by the Gaussian mixture model for separating materials \cite{huang2008new}. Another recent trend is to use deep artificial neural networks for image processing and segmentation \cite{egmont2002image}. 
Deep Convolutional Neural Networks (CNNs) are developing actively due to the increase in computational performance of graphical accelerators (GPUs). In segmenting SEM images they produced results of superior quality compared to conventional algorithms \cite{ciresan2012deep}. Further segmentation improvements were demonstrated with the architecture of fully-convolutional neural networks (FCNN) \cite{long2015fully}. One of the most popular FCNN architectures is the well-known U-net \cite{ronneberger2015u}, which was also extended to 3D \cite{cciccek20163d}. Unlike previously FCNN models, the U-net architecture allows training on a small dataset and it is widely used for solving quantification tasks in various scientific studies \cite{falk2019u}. 

In this paper, we develop a new method for automatic segmentation of dynamic CT images during the methane-hydrate formation in sandy samples. The approach consists of two sequential steps: (1) applying the U-net neural network segmentation model, (2) statistical clustering based on the Gaussian mixture model. The method allows for conducting accurate segmentation of 3D data with the presence of low-contrast phases caused by the methane hydrate formation in brine saturated sandy samples. The efficacy of the proposed method is demonstrated by an automatic quantitative analysis of CT data from the tomographic experiment described in \cite{nikitin2020dynamic}. The paper is organized as follows. 
In Section \ref{sec2} we briefly describe the experimental setup, data acquisition, and data processing procedures to generate 3D volumes for further segmentation.
In Section \ref{sec3} we introduce our two-step automatic segmentation approach involving the usage of U-net architectures (2D and 3D) and the Gaussian mixture model.
Section \ref{sec4} presents a quantitative analysis of the segmentation results and discuss new insights about the the formation process. Conclusions and outlook for further gas-hydrates studies are given in Section \ref{sec5}.

\section {Data description}\label{sec2}

Tomographic data analyzed in this study were acquired during a dynamic in-situ experiment at the bending magnet beamline 2-BM of the Advanced Photon Source, Argonne National Laboratory, see \cite{nikitin2020dynamic} for details. For gas-hydrate formation we used an environmental cell filled with silica sand and water, supplied with methane gas under high pressure, and continuously cooled at -7~$^\circ$C. The cell was rotated and scanned with a parallel X-ray beam every 15 min to capture the sample states without gas hydrates and the states where the hydrate is continuously forming by filling the pore space. Acquisition of one tomographic dataset corresponding to a 180 degrees sample rotation took 70 sec.  

For obtaining qualitative tomographic reconstruction we used TomoPy~\cite{gursoy2014tomopy} package with the command line interface \textit{tomopy-cli}\footnote{\url{https://tomopycli.readthedocs.io/en/latest/}} for organizing a reconstruction pipeline. The reconstruction pipeline included common processing functions from TomoPy, such as ring removal, phase retrieval filtering, and filtered backprojection implemented via the log-polar-based method~\cite{andersson2016fast}.  

Fast data collection in the dynamic synchrotron experiment on the gas-hydrate formation produces more than 200~GB of reconstructed 3D images per experimental day. Each reconstructed 3D image volume has $1224\times1224\times512$ size in 32-bit precision, and corresponds to a $4.3\times4.3\times1.8$ mm$^3$ real sample volume with  $3.5\mu m^3$ voxel size. Each 3D image is reconstructed from 1500 8-bit projections of size $1224\times512$. Reconstruction of one dataset takes about 10 min on an Intel Xeon-type processor. In this work, for the development and testing of our segmentation algorithm we used 66 tomographic images at different times during the hydrate formation. 

According to the presence and quantity of each phase (sand grains, methane hydrate, NaBr brine, and methane gas) during the formation process, we have distinguished three main experimental stages: 1) before the methane hydrate formation, 2) during the methane hydrate formation, 3) after the methane hydrate formation. At the first stage, tomographic images contain methane gas, sand grains, and NaBr brine. The second stage covers the sample states with methane gas, sand grains, methane hydrate, NaBr brine, and a mixture of NaBr brine and methane hydrate. The third stage corresponds to images with methane gas, sand grains, and methane hydrate. 

\begin{figure*}
  \begin{minipage}[c]{0.7\textwidth}
    \includegraphics[width=0.95\textwidth]{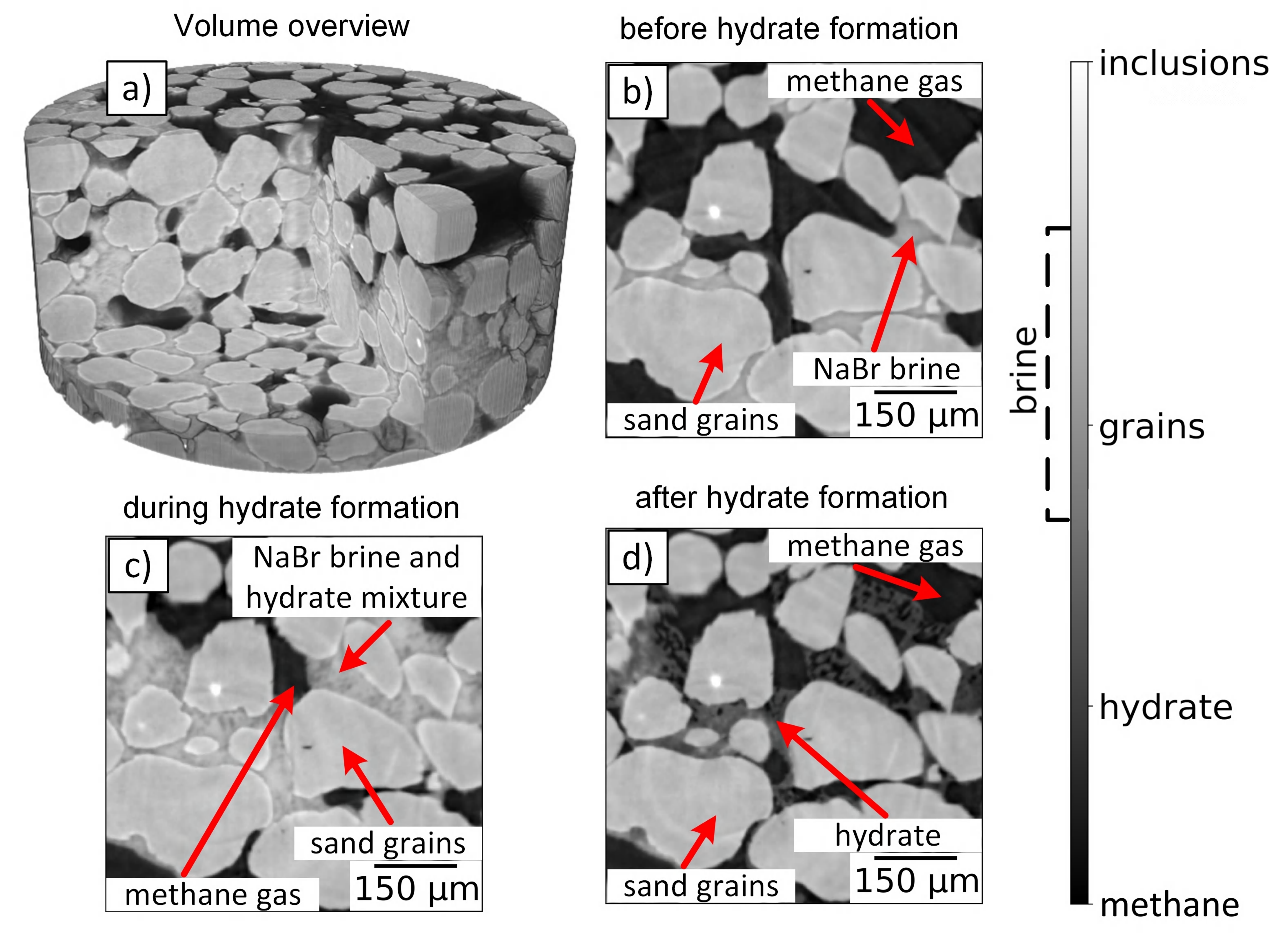}
  \end{minipage}\hfill
  \begin{minipage}[c]{0.3\textwidth}
    \caption{An example of 3D reconstruction of the hydrate containing sandy sample (a). Cropped parts of the slices showing a typical content of tomographic data obtained at the main stages of the tomographic experiment (b,c,d).
    } \label{fig:data_description}
  \end{minipage}
\end{figure*}

An example of reconstructed volume after a cylindrical cut is shown in Figure \ref{fig:data_description}, a. Panels b,c,d of the same figure show three examples of cropped slices obtained at the different experimental stages. The figure is equipped with notes indicating different materials: black color corresponds to the methane gas, dark gray -- to the gas hydrate, light gray -- to salty water (NaBr brine), very light gray -- to sand grains. One can also observe regions with a mixture of the NaBr brine and the hydrate.

The use of salt brine is common in micro-computed tomography imaging of methane gas hydrates because the brine acts as an increasing phase contrast agent to separate the hydrate and water in images. \cite{lei2018laboratory} demonstrated efficient phase contrast-enhancing methods based on using NaBr, KI and NaCl brines as hydrate-forming fluids, as opposed to the regular formation with distilled water. These methods also simulate the natural scenario of hydrate formation - formation in the bottom sediments of the seas where water has some salinity. 

\begin{figure*}
  \begin{minipage}[c]{0.7\textwidth}
    \includegraphics[width=0.95\textwidth]{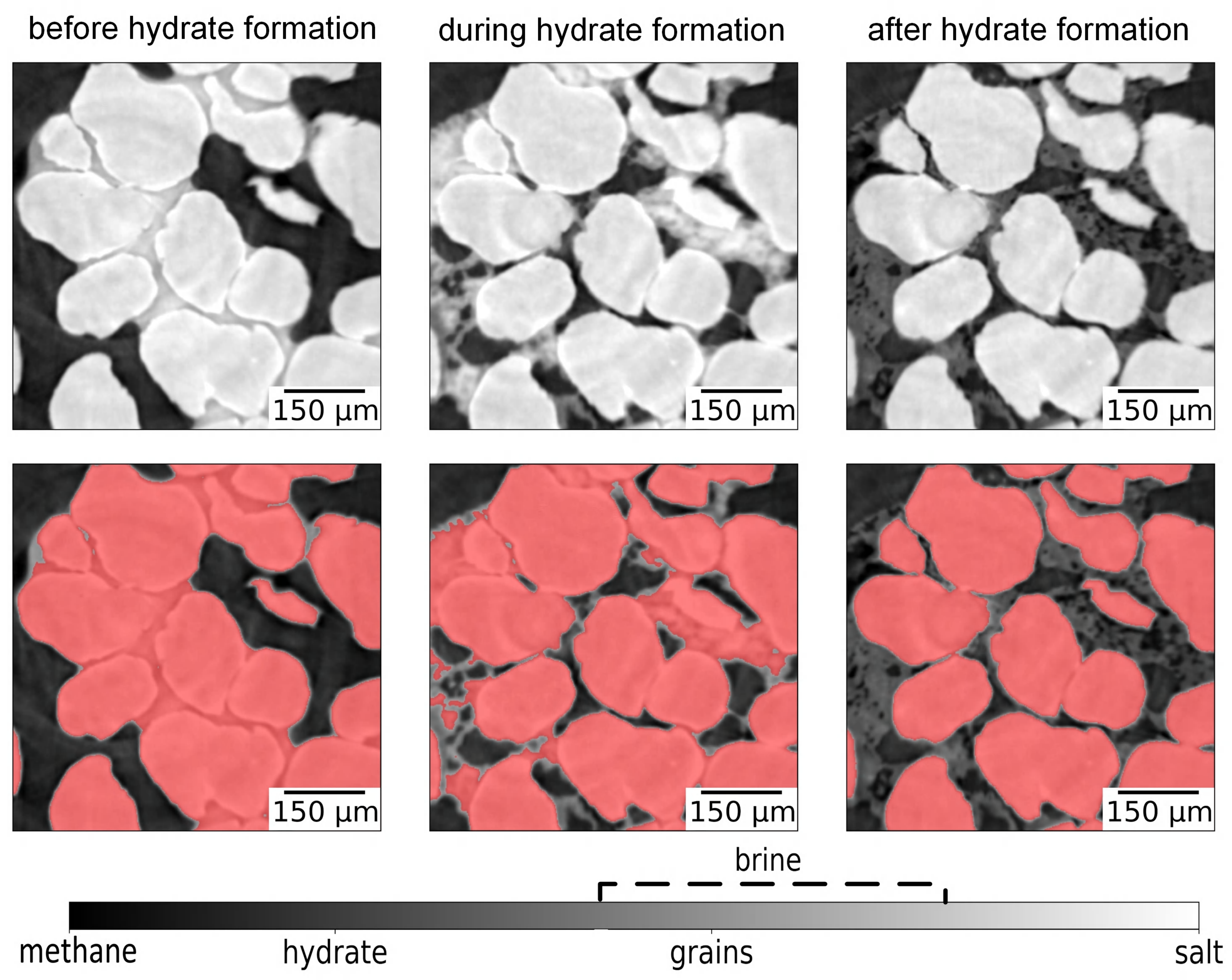}
  \end{minipage}\hfill
  \begin{minipage}[c]{0.3\textwidth}
    \caption{Failing segmentation of sand grains by using a general threshold-type method due to low contrast between the grain phase and other phases inside the pore space.
    } \label{fig:low_phase}
  \end{minipage}
\end{figure*}

In this work we tested different brine salinity levels and chose 10\% NaBr as it demonstrates a more favorable hydrate-water contrast for further segmentation procedures. However, the increase of contrast between one pair of phases reduces the contrast between other pairs of phases. Specifically, phase contrast between water and sand grains is significantly reduced with the brine appearance. Conventional segmentation algorithms based on gray level separation become ineffective in this case. Figure~\ref{fig:low_phase} shows an example of failing segmentation of the sand grains using the thresholding algorithm. The top row shows different sample slices containing grains and NaBr brine, and the bottom row demonstrates the segmentation results not allowing to separate two phases.

\section{Segmentation procedure}
\label{sec3}

In this section we present basic concepts and implementation details of the two-steps segmentation technique developed for segmenting dynamic CT data obtained during the methane hydrate formation in sandy samples. The proposed segmentation procedure involves application of the following methods: 

\begin{enumerate}
  \setlength\itemsep{0.01cm}
  \item Sand grains segmentation using the U-net model.
  \item Pore space phases segmentation using a clustering algorithm based on the Gaussian mixture model.
\end{enumerate}

The first step is needed to separate sand grains and brine filled phases (NaBr brine and the mixture of NaBr brine and methane hydrate). For this, we apply semantic segmentation of grains by using models based on deep fully-convolutional U-net networks \cite{ronneberger2015u}. At the second step, the segmented areas of sand grains are subtracted from the images, and separation of the remaining phases is carried out with the clustering algorithm based on the Gaussian mixture model \cite{huang2008new}. 
 The whole segmentation procedure was implemented in Python using Tensorflow and scikit-learn packages.

\subsection{U-net based models for sand grains segmentation}

For the grains segmentation we used both 2D and 3D U-net implementations based on the models described in \cite{ronneberger2015u, cciccek20163d}. The architecture of the {U-net} models can be divided into the encoder and the decoder parts. The encoder consists of the convolutional and downsampling layers applied step by step to compute feature maps of the input data. The decoder takes low-resolution feature representations and generates the mask using the transposed convolutional and convolutional layers. The resulted masks have the same size as the input data. 

\begin{figure*}
  \begin{minipage}[c]{0.7\textwidth}
    \includegraphics[width=0.95\textwidth]{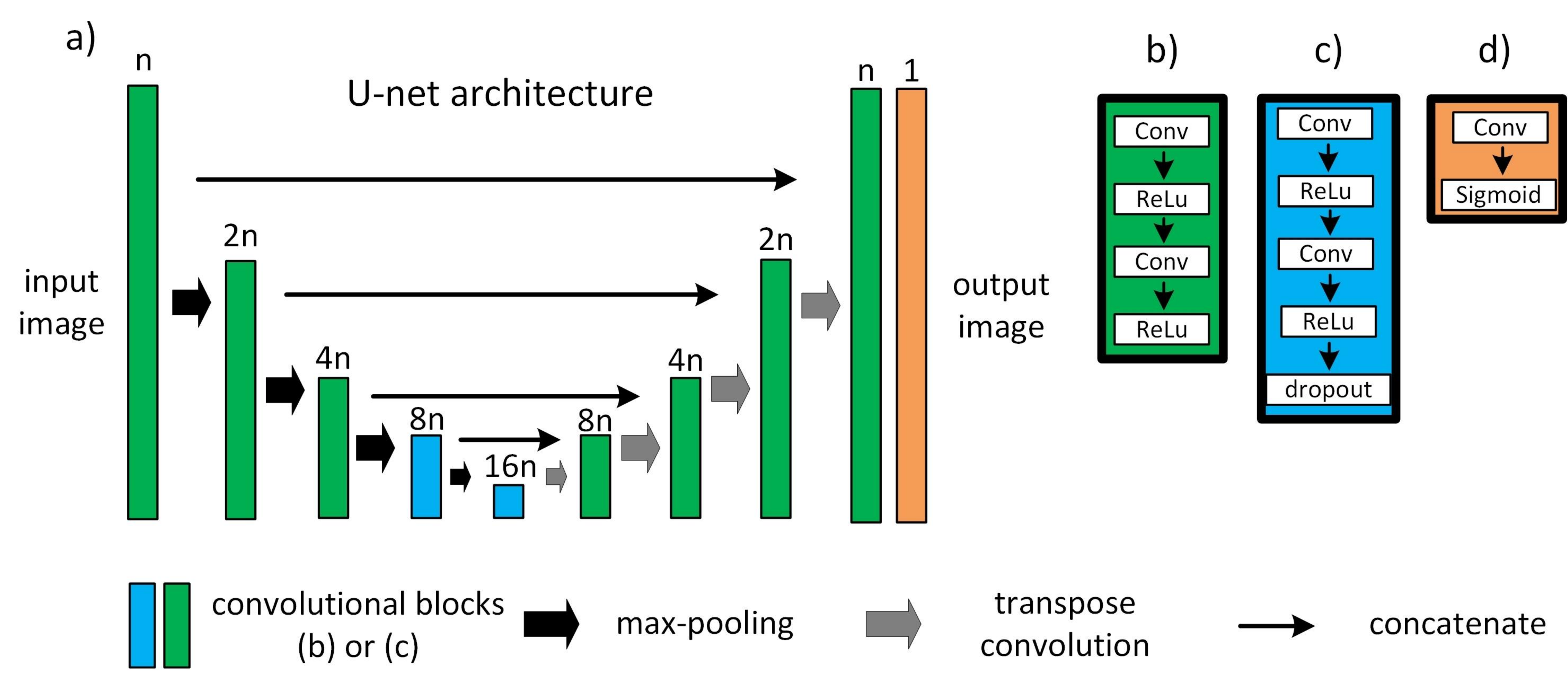}
  \end{minipage}\hfill
  \begin{minipage}[c]{0.3\textwidth}
    \caption{Schematic overview of the 2D and 3D U-net architectures used in this work. Colored squares denote the convolutional blocks and the arrows denote different math operations.
    } \label{fig:U-net}
  \end{minipage}
\end{figure*}

A detailed description of the proposed U-net architecture is presented in Figure \ref{fig:U-net}. One can see that the encoder and the decoder parts consist of 4 convolutional blocks (Figure \ref{fig:U-net}a). Each block is highlighted by the color depending on the layers it contains. So, we used an architecture with two types of convolutional blocks and one final convolutional layer. Green color indicates the blocks consisting of two sequential convolutional layers with the $ReLu$ activation function (Figure \ref{fig:U-net}b). Blue color indicates the blocks with two convolutional layers followed by dropout (Figure \ref{fig:U-net}c). Orange color indicates the final convolutional layer with the kernel size of 1 pixel and the sigmoid activation function (Figure \ref{fig:U-net}d). 

We implemented the described architecture in both 2D and 3D. The number of filters for the first convolutional block was 32 and 16 for the 2D and 3D models, respectively. 

\subsubsection{Preparation of training and validation datasets}

The supervised learning strategy we propose requires labeled datasets for training. Manual labeling of tomographic images is time-consuming, especially in 3D, therefore, in this work we avoided manual data labeling following the strategy described below. 

\begin{figure*}
    \centering
     \includegraphics[width=0.9\textwidth]{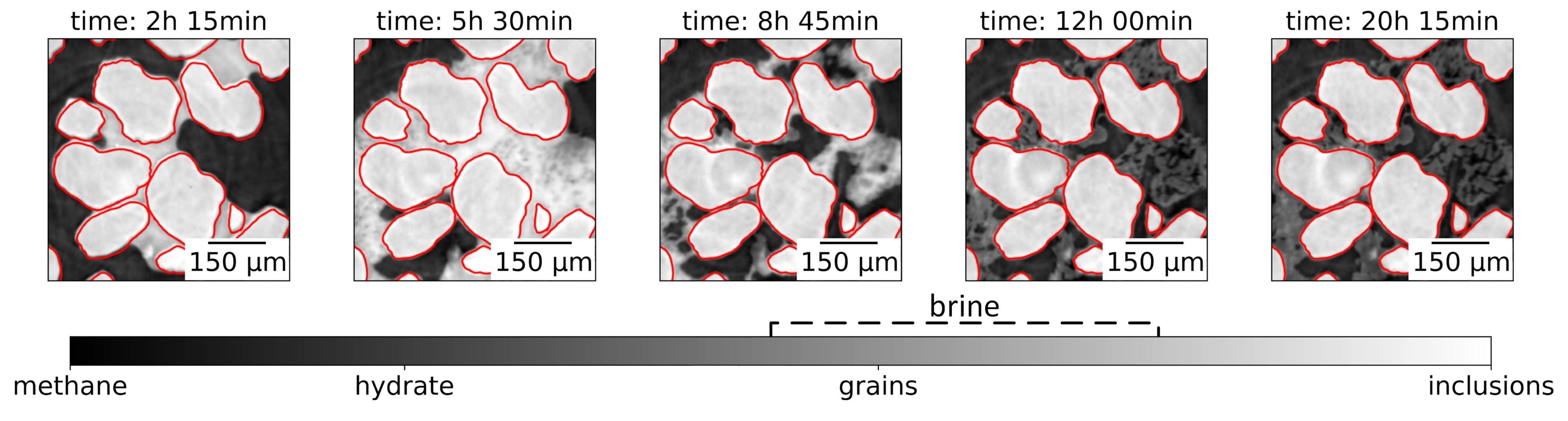}
    \caption{Demonstration that grains are motionless during the hydrate formation process: the red curves indicating the borders of the grains at the beginning of the experiment have the same positions in all images.}
    \label{fig:grains_motionless}
\end{figure*}

To automate the data labeling procedure, we rely on one of the main properties of the hydrate formation in porous media: sand grains remain motionless while pore brine is mobile and almost disappears by the end of the experiment due to converting into the gas hydrate, see \cite{nikitin2020dynamic} for details. As mentioned earlier, there are mostly three phases presented in the CT images at the end of the hydrate formation: methane gas, methane gas hydrate, sand grains. Since these phases are well-separated in the gray-level images, it is possible to automatically segment the grains by conventional thresholding algorithms, see the right panel in Figure \ref{fig:grains_motionless}. Given that the grains are motionless during the experiment, the resulting grain masks can be applied to images from previous time steps. Thus we just have to choose regions and times when there was brine in the pore space, see the first two panels in Figure \ref{fig:grains_motionless}.  

Following this automated labeling approach we prepared a labeled dataset consisting of 1134 sub-volumes ($256\times256\times256$) overlapping by not more than 145 pixels. 80 \% of this dataset was used for training and validation. The remaining 20 \% was used for testing. The training dataset for 2D segmentation was prepared by slicing the 3D sub-volumes.

\subsubsection{Training results and quality comparison}

The 2D and 3D segmentation models were trained using the same deep-learning training strategy. For explanation of all ML terms and techniques used in this section we refer to \cite{goodfellow2016deep}. The training procedure involved the \emph{binary\ cross-entropy} loss function minimization with the \emph{Adam} optimizer and \emph{accuracy} metric. The metric and loss function values on the training and validation sets were monitored during training to avoid the model overfitting and determine the optimal number of training epochs. As a result, we chose the number of epochs equal to 15. Segmentation quality on the test dataset was measured by the \emph{mIoU} metric. To analyze the performance of 2D and 3D U-nets for processing large datasets we measured the computational time for the training and prediction stages. 

\begin{table*}
\begin{center}
\begin{minipage}{\textwidth}
\caption{Quality and performance testing of 2D and 3D U-net models for segmenting sand grains in $1224\times1224\times512$ reconstructions of hydrate-bearing samples}\label{tab:Table1}%
\begin{tabular}{@{}llll@{}}
\toprule
NN architecture& prediction time (one volume)  & training time (15 epochs) & mIoU\\
\midrule
2D U-net    & 19 sec   & 5.4 min  & 0.931  \\
3D U-net    & 14 sec   & 102 min  & 0.976  \\
\botrule
\end{tabular}
\end{minipage}
\end{center}
\end{table*}

Segmentation quality and performance results are shown in Table~\ref{tab:Table1}. For estimating the training time (for 15 epochs) we have averaged 10 independent training runs on the same dataset. The prediction time corresponds to the time needed for segmenting a $1224\times1224\times512$ image volume divided into 75 patches of size $256\times256\times256$ and with overlapping by 14 pixels. The training and testing procedures were implemented in Python using TensorFlow 2.7.0 package. The performance testing was run on an NVidia Tesla V100 graphics accelerator with 16 GB of memory.  
Note that the 3D model is 5 seconds faster than the 2D model in prediction (segmentation). However, the 2D model is 20 times faster in training. The \emph{mIoU} metric shows that the 3D model demonstrates $4\%$ better segmentation quality compared to the 2D model.

\begin{figure*}
    \centering
    \includegraphics[width=0.75\textwidth]{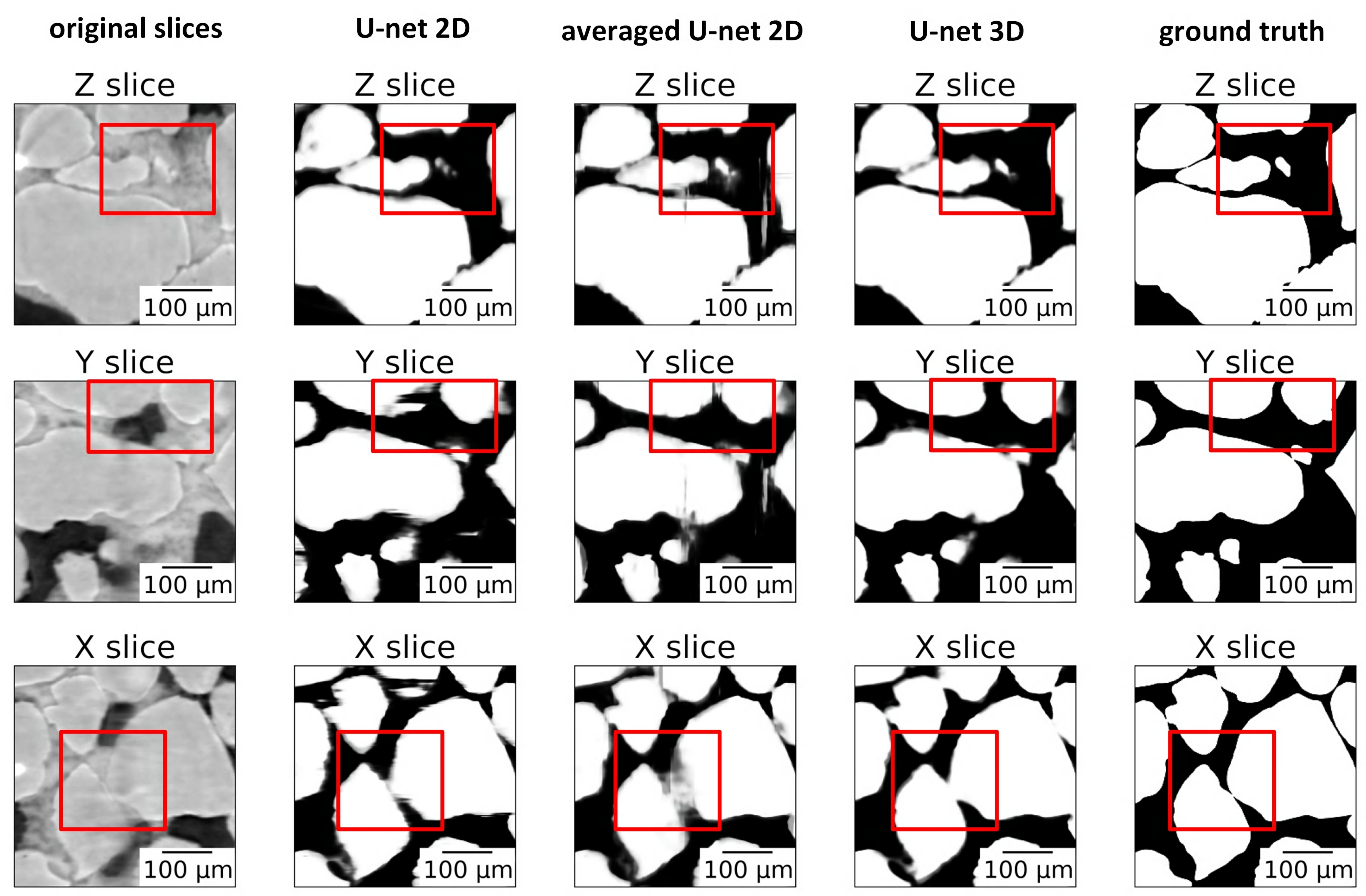}
    \caption{Comparison of segmentation results by the U-net 2D, averaged U-net 2D, and U-net 3D neural networks applied to a sub-volume from the test sample. Each panel shows X, Y and Z central slices from the sub-volume. Red rectangles correspond to the regions with significant differences in segmentation quality.}
    \label{fig:results_compare}
\end{figure*}

Figure \ref{fig:results_compare} demonstrates a comparison of the grains segmentation quality for the 3D and 2D U-net models. Each row of the figure shows the central slice of a sub-volume from the test dataset: horizontal slice in the upper row (labeled as Z-slice), vertical slice orthogonal to Y axis in the middle row (labeled as Y-slice), vertical slice orthogonal to X axis in the bottom row (labeled as X-slice). One can see the original slices in the leftmost column and the labeled slices (ground truth) in the rightmost column. Three columns in the middle show segmentation results of applying different U-net models: 2D U-net applied to horizontal slices in a slice-by-slice manner (2nd column); 2D U-nets applied in a slice-by-slice manner in directions orthogonal to axes X, Y, and Z, respectively, followed by averaging the results (3rd column); 3D U-net applied to the whole volume (4th column). 

Note that 2D segmentation (2nd and 3rd columns in Figure~\ref{fig:results_compare}) results in `comb'-type artifacts caused by the fact that the 2D model does not take into account information from neighboring slices, and may easily build non-smooth boundaries in the direction of the slice sliding. 
The areas with the strongest artifacts are highlighted by the red squares. 
On the contrary,  the 3D model is able to build smooth boundaries in all directions and provides the best segmentation result. 
Our conclusions are inline with the assumption in \cite{deniz2018segmentation} that the end-to-end 3D CNN segmentation model is more favorable for segmenting challenging 3D images. 

\subsection{Clustering with the Gaussian mixture model}

After extracting segmented sand grains from the original image, the remaining phases (gas, hydrate, brine and hydrate mixture, brine) were separated by the global threshold method. The main difficulty in determining the global threshold was that some of the remaining phases have unstable in time gray level ranges. These instabilities were mainly due to the effect of salt ions exclusion from the water consumed for the hydrate formation process \cite{chen2020pore}. Thus the hydrate formation slowly increases the brine salinity in time, shifting the gray-level distribution between phases. As a result, we need to determine the gray-level thresholding ranges at each time step.

In this work, we used a clustering algorithm based on the Gaussian mixture model (GMM) to automatically determine the global threshold at each time step. This model allows decomposing the mixed normal distribution into the sum of Gaussians, followed by iterative calculating their parameters with the Expectation Maximization (EM) algorithm \cite{balafar2014gaussian}. Definition of the starting GMM requires setting the number of mixture components and initializing parameters for Gaussian distributions. We defined the number of components equal to 4 according to the maximum number of phases that can be presented in our CT data. To include the time dependence of the data, we used the means, covariances, and weights calculated at the previous time step as parameters for initializing the starting model at the current time step. For the first time step, the model was initialized using the k-means algorithm.

\begin{figure*}
    \centering
    \includegraphics[width=1\textwidth]{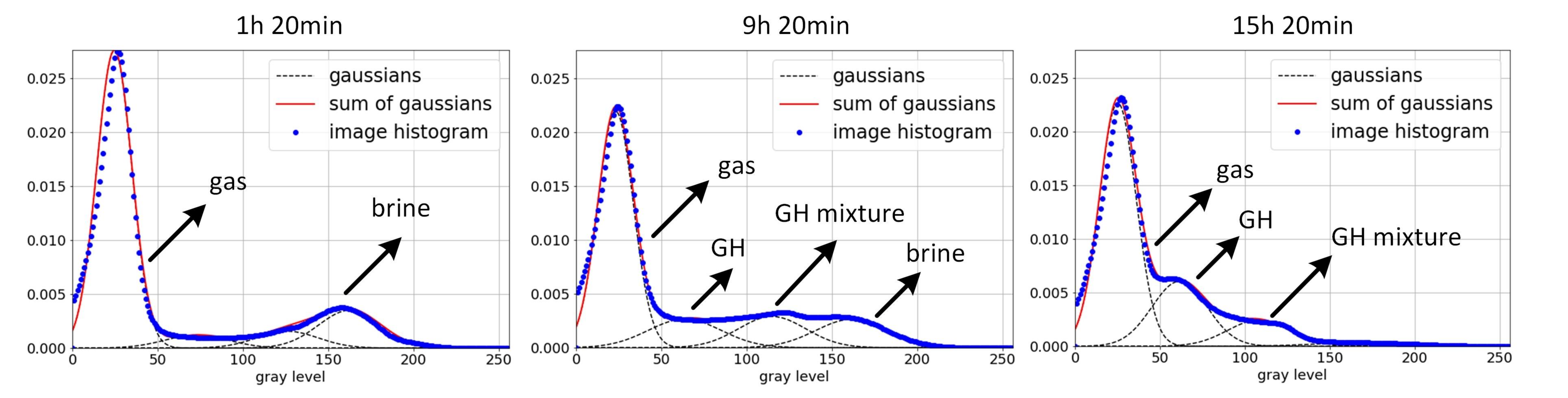}
    \caption{Histogram approximation by the Gaussian mixture model. Each graph corresponds to CT data associated to different times in the hydrate formation experiment. Data histogram is shown with blue dots, Gaussians fitted with GMM - black dots, sum of the Gaussians - red line.}
    \label{fig:hist_decompose}
\end{figure*}

Figure \ref{fig:hist_decompose} shows the results of the mixture distribution decomposition using the approach described above. Three time steps were chosen as examples describing the main stages of the experiment: before hydrate formation (1 h 20 min), during hydrate formation (9 h 20 min), and after hydrate formation (15 h 20 min). The experiment time is measured from the moment when P-T conditions for the hydrate formation became stable. Each panel in Figure \ref{fig:hist_decompose} includes experimental and predicted probability density functions (PDF) marked by blue and red colors, respectively. Gaussian distributions for each phase from the GMM model are marked by the black dashed line. The experimental PDF is calculated from the histogram of the images, while the predicted PDF is calculated as a sum of predicted Gaussian distributions. Black arrows indicate phases corresponding to each Gaussian. One can see that the proposed mixed distribution decomposition into 4 components gives a good agreement between the experimental and predicted PDF. These 4 components were associated with the following phases: sand grains, methane gas (gas), methane gas hydrate (GH), methane hydrate and NaBr brine mixture (GH mixture), NaBr mixture.     

\begin{figure*}
  \begin{minipage}[c]{0.7\textwidth}
    \includegraphics[width=0.95\textwidth]{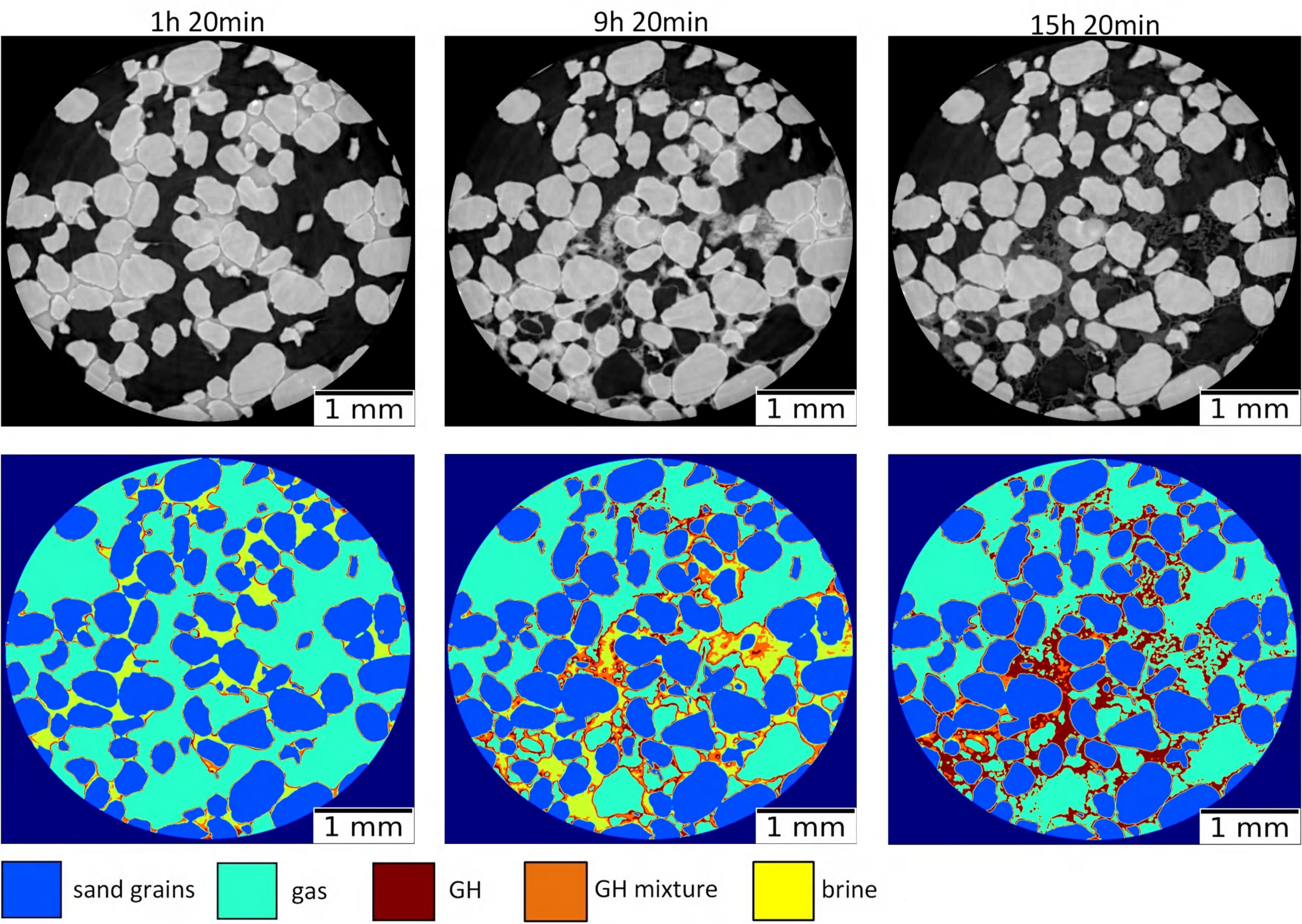}
  \end{minipage}\hfill
  \begin{minipage}[c]{0.3\textwidth}
    \caption{Extracted horizontal slices from the clustered 3D sample volumes by the proposed two-step segmentation model. Each panel corresponds to CT data from different experiment times.
    } \label{fig:GMM_results}
  \end{minipage}
\end{figure*}

Figure \ref{fig:GMM_results} shows the results of clustering CT images with the proposed two-step segmentation algorithm. The top row shows central slices of 3D image sample volumes at different experiment times. Histogram approximation plots in Figure \ref{fig:hist_decompose} were prepared with making use of these data. After decomposing the gray-level curve into separate Gaussians one can use thresholding for separating corresponding phases. Thresholding boundaries were chosen as intersections between the Gaussians and yielded segmentation masks shown in the bottom row of Figure \ref{fig:GMM_results}. Colors correspond to different phases in the GMM: sand grains (blue), gas (green), gas hydrate (brown), mixture of brine and hydrate (orange), pure brine (yellow).

\section{Results and discussion}
\label{sec4}

\begin{figure*}
  \begin{minipage}[c]{0.7\textwidth}
    \includegraphics[width=0.95\textwidth]{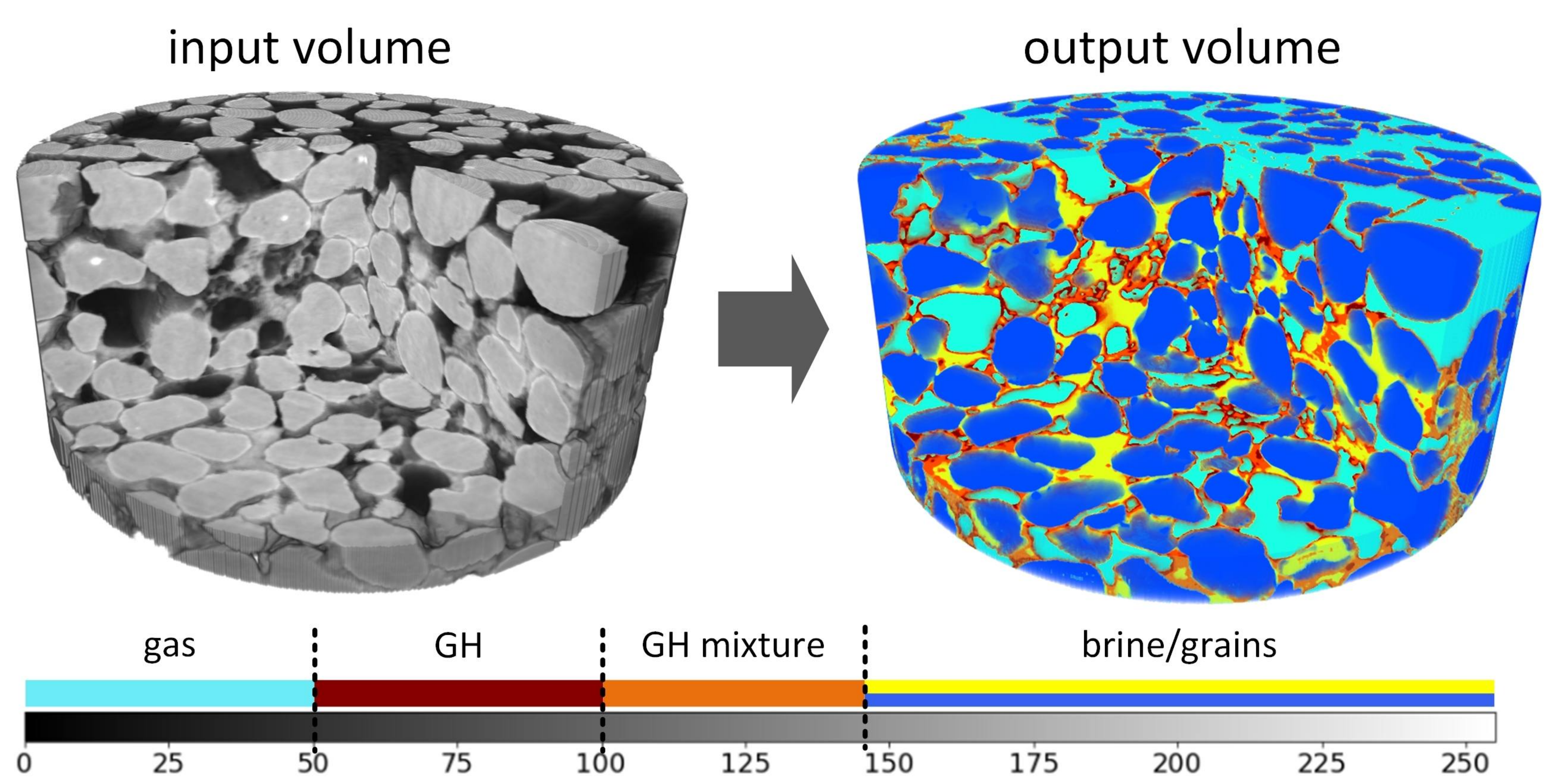}
  \end{minipage}\hfill
  \begin{minipage}[c]{0.3\textwidth}
    \caption{Example of applying the proposed segmentation method in 3D (color code is the same as in Figure~\ref{fig:GMM_results}).
    } \label{fig:final_example}
  \end{minipage}
\end{figure*}

Let us discuss some applications of the proposed segmentation method. 
First, we can use the developed two-step segmentation algorithm to process 3D CT image volumes in an automatic manner, where each CT volume is provided as input and its automatic segmentation is formed as output, see Figure \ref{fig:final_example}. The colors of the materials here are the same as in Figure \ref{fig:GMM_results}: sand grains (blue), gas (green), gas hydrate (brown), pure brine (yellow), mixture of brine and gas hydrate (orange). Note that the solid sample matrix consists of two materials: sand grains (invariant) and gas hydrate (changing in time). Thus, as the first applications of our segmentation we can consider the monitoring of the 3D pore space geometry changing in time. The resulting time-resolved 3D models can be further used in Digital Rock Physics simulations for estimating changes in the sample permeability and in other petrophysical properties during the hydrate formation. 

Another application is related to the quantitative estimation of the sample material changes in time. Pore-space saturation of each particular phase can be computed as a sum of all voxels corresponding to this phase, divided by the sum of voxels corresponding to the entire pore space. For example, methane gas saturation in pores of the volume in Figure \ref{fig:data_description} is 61.8 \%. The situation is more complicated for such materials as the pore brine or gas hydrate. Based on our Gaussian mixture model, the gas hydrate is presented in two phases: as the GH phase (brown color in Figure~\ref{fig:GMM_results}) and as a part of the GH mixture (orange color in Figure~\ref{fig:GMM_results}). In the latter case, voxels should be decomposed into relative amounts of brine and hydrate. Here we assume a linear dependence of the relative content of these phases on the gray level within the interval of the GH mixture phase. Therefore, for quantitative estimation of the gas-hydrate material in a 3D volume we compute the fraction of the GH-mixture voxels following our linear dependence assumption and add it to the number of GH phase voxels. Similar estimations can be done for the pore brine material. To compute the gas hydrate and brine saturation, the obtained values are divided by the number of voxels representing the whole pore-space volume. 

\begin{figure*}
  \begin{minipage}[c]{0.7\textwidth}
    \includegraphics[width=0.95\textwidth]{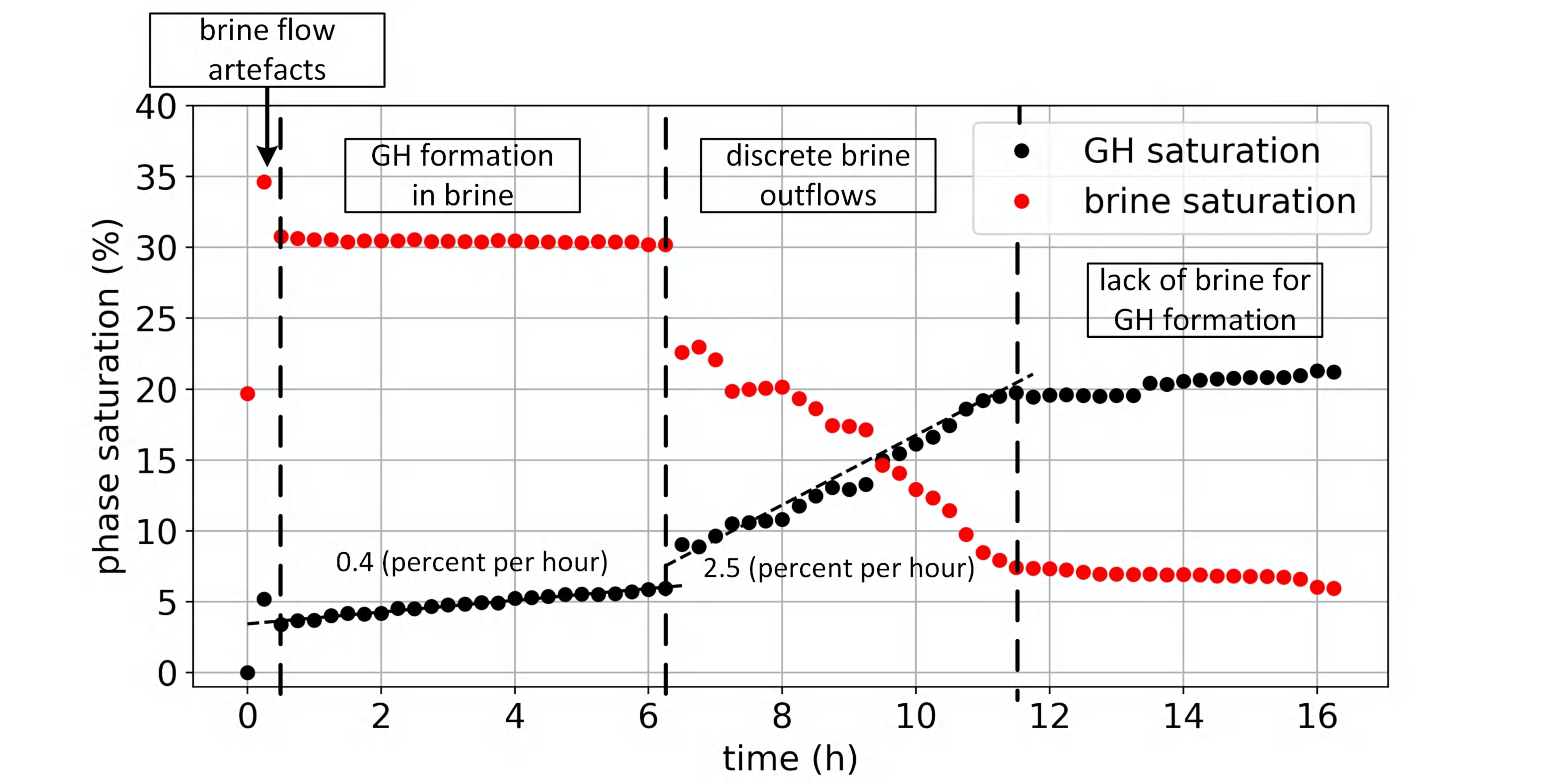}
  \end{minipage}\hfill
  \begin{minipage}[c]{0.3\textwidth}
    \caption{Hydrate and water saturation plots over the whole hydrate growth process. The red curve corresponds to water saturation, black curve - to hydrate saturation.
    } \label{fig:params_calc}
  \end{minipage}
\end{figure*}

In Figure \ref{fig:params_calc} we show the saturation of the brine (red curve) and hydrate (black curve) at different times during the experiment. These automatically generated curves give new insights into the hydrate formation process. First, we see two distinct periods of the hydrate growth. Gas-hydrate growth is slow during the first 6 hours. Close examination of the images revealed that the hydrate growth rate changed after a massive volume of brine has moved away out of the scanned region shortly after 6 hours. 
Apparently, the gas transfer into massive water volumes is slow resulting in a slower hydrate-formation rate of 0.4 \% per hour, see the interval 0 -- 6 hours in Figure \ref{fig:params_calc}. After massive water outflow the surface of the brine-gas contact increases resulting in a faster gas-hydrate formation rate of 2.5 \% per hour, see the interval 6.5 -- 11 hours in Figure \ref{fig:params_calc}. 
Finally, the hydrate growth reaches saturation after 11 hours of the experiment.

\section{Conclusions}
\label{sec5}

With the development of new powerful synchrotron sources, data acquisition rates will become significantly higher, making it possible for X-ray beamline instruments to generate $100\times$ more data per experiment~\cite{chenevier2018esrf,fornek2019advanced}.
Large data volumes require automatic segmentation that is particularly challenging for imaging gas-hydrate formation in porous samples. The phases have weak gray-level contrast (grains and brine) that is also evolving in time (brine salinity increases during the hydrate formation). We proposed a two-step segmentation algorithm. First, 3D U-net architecture is used for the most challenging problem of separating the sand grains. Then we segment the remaining phases using the Gaussian mixture model to adapt the global thresholding levels that change in time. The source code is publicly available, see 'Code availability section' below.

We showed application of the proposed hybrid machine learning method for quantitative estimation of the hydrate-saturation changes in time. The hydrate-saturation curve revealed two distinct periods of the hydrate growth: slow hydrate growth in the beginning, followed by a five-time faster hydrate growth after the massive water outflow. These observations are important for a better understanding of the kinetics of the hydrate formation in porous medium -- massive water outflows in pores improve brine-gas contact facilitating the hydrate formation. 

The proposed method may become an essential tool for imaging and analyzing gas hydrates in high temporal and spatial resolution. 
In fast-evolving dynamic systems, automatic segmentation, classification, and detection may allow for steering tomographic experiments, e.g. changing environmental conditions (pressure, temperature, electrical charge) based on real-time sample states. AI-based steering techniques will play a very important role in future complex dynamic experiments at brilliant light sources.

\section*{Declarations}

\bmhead{Funding}
This research used resources of the Advanced Photon Source, a U.S. Department of Energy (DOE) Office of Science User Facility operated for the DOE Office of Science by Argonne National Laboratory under Contract No. DE-AC02-06CH11357. The research was partly supported by Russian Ministry of Science and Higher Education under project No. FWZZ-2022-0030.

\bmhead{Conflict of interests}
The authors declare that they have no conflict of interests.

\bmhead{Data availability}
Data underlying the results presented in this paper are available in the following Ref. \cite{Fokin2022}


\bibliography{bibliography}


\end{document}